\documentclass[aps,prl,reprint,groupedaddress,floatfix]{revtex4-1}
\usepackage{graphicx}
\usepackage{hyperref}
\usepackage{natbib}
\usepackage{amsmath}
\usepackage{amsfonts}
\usepackage{amssymb}
\usepackage{bm}
\begin{document}

\title{Cluster-induced Deagglomeration in Dilute Gravity-driven Gas-solid Flows of Cohesive Grains}

\author{Peiyuan Liu}
\author{Christine M. Hrenya}
\email[Corresponding author: ]{hrenya@colorado.edu}
\affiliation{Department of Chemical and Biological Engineering, University of Colorado Boulder, Boulder, Colorado, 80309, USA}

\date{\today}

\begin{abstract}
Clustering is often presumed to lead to enhanced agglomeration between cohesive grains due to the reduced relative velocities of particles within a cluster. Our discrete-particle simulations on gravity-driven, gas-solid flows of cohesive grains exhibit the opposite trend, revealing a new mechanism we coin ``cluster-induced deagglomeration.'' Specifically, we examine relatively dilute gas-solid flows, and isolate agglomerates of cohesive origin from overall heterogeneities in the system – i.e., agglomerates of cohesive origin and clusters of hydrodynamic origin. We observe enhanced clustering with an increasing system size (as is the norm for noncohesive systems) as well as reduced agglomeration. The reduced agglomeration is traced to the increased collisional impact velocities of particles at the surface of a cluster – i.e., higher levels of clustering lead to larger relative velocities between the clustered and nonclustered regions, thereby serving as an additional source of granular temperature. This physical picture is further evidenced by a theoretical model based on a balance between the generation and breakage rates of agglomerates. Finally, cluster-induced deagglomeration also provides an explanation for a surprising saturation of agglomeration levels in gravity-driven, gas-solid systems with increasing levels of cohesion, as opposed to the monotonically increasing behavior seen in free-evolving or driven granular systems in the absence of gravity. Namely, higher cohesion leads to more energy dissipation, which is associated with competing effects: enhanced agglomeration and enhanced clustering, the latter of which results in more cluster-induced deagglomeration.
\end{abstract}

\maketitle

Due to the dissipative nature of solid-solid and fluid-solid interactions, granular and gas-solid flows develop hydrodynamic instabilities that lead to \textit{clusters}: local regions of high solid concentration  \cite{Goldhirsch1993, Glasser1998, Wylie2000, Agrawal2001, Brilliantov2004, Aranson2006, Mitrano2013, Yin2013, Fullmer2017}, which are absent in molecular fluids. Inter-particle cohesion, such as van der Waals force \cite{Castellanos2005}, liquid-bridging \cite{Herminghaus2005, Mitarai2006} and electrostatics \cite{Pahtz2010, Lee2015}, enhances energy dissipation during particle collisions \cite{Thornton1991, Gollwitzer2012}, causing the formation of \textit{agglomerates} \cite{Royer2009, Waitukaitis2011}. Unlike the loose collection of particles characterizing clusters, agglomerates refer to particles held together in enduring contacts by cohesion \cite{Horio1992}. Both clustering and agglomeration significantly impact reaction rates, momentum, heat and mass transfer in multiphase flows \cite{Fullmer2017}.

Here, we aim to understand the interplay between clusters and agglomerates. For granular systems (no fluid) without gravity, the formation of clusters enhances agglomeration. Namely, in both free-evolving \cite{Ulrich2009, Murphy2015, Singh2018} and driven \cite{Blair2003no2, Weber2004, Takada2014} systems, particles in clusters have higher collision frequency due to the increased local number density. Therefore, the collisional impact velocities (relative particle velocities prior to collisions) of particles in clusters decay faster than particles in the surrounding, less-dense regions \cite{Losert1999, Goldhirsch2003}. With reduced impact velocities, particles are more likely to agglomerate upon collision \cite{Kantak2004, Donahue2010}. Moreover, the rapid energy dissipation within clusters results in a pressure gradient across the cluster interface which promotes the migration of free particles towards clusters \cite{Goldhirsch1993, Miller2004}, further increasing the cluster size and thus the possibility of agglomeration.

In this Letter, we study the relationship between clustering and agglomeration in dilute gas-solid flows of lightly cohesive particles in unbounded fluidization via discrete-particle simulations. Unlike granular systems under zero gravity, gas-solid flows are driven by gravity and have two additional sources of clustering beyond dissipative particle collisions \cite{Fullmer2017}: relative motion between gas and solid phases (mean drag) \cite{Agrawal2001} and dissipation of granular energy due to gas viscosity (thermal drag) \cite{Wylie2000}. We report an unexpected response of agglomerates to increasing system size. Namely, analogous to fluid turbulence, the level of clustering increases with system size, as is also observed in non-cohesive systems \cite{Capecelatro2016}. However, unlike granular systems, the clustering in gas-solid systems does not enhance agglomeration; instead, the degree of agglomeration reduces with increased system size. This observation is surprising since particles within a cluster are characterized by reduced impact velocities, which favor enhanced agglomeration. Based on an analysis of particle velocities, we uncover the physical mechanism for this surprising behavior - \textit{cluster-induced deagglomeration} - and establish an analytical model to predict the resulting degree of agglomeration. We then demonstrate the robustness of the mechanism at higher cohesion levels.

\begin{figure}[t]
 \includegraphics[width=0.9\columnwidth]{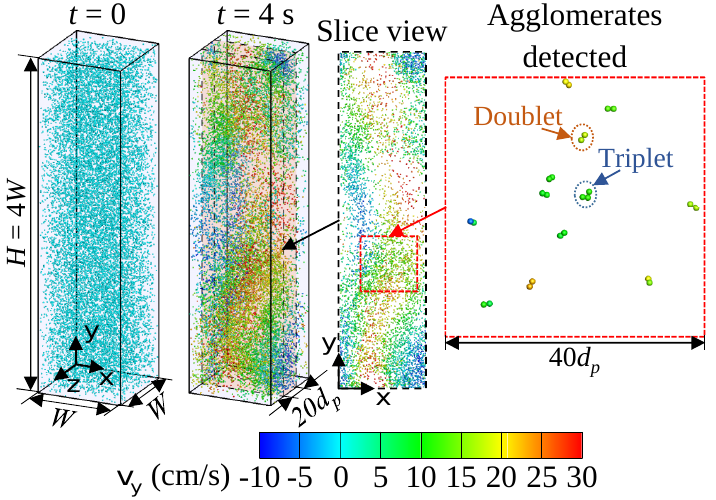}
 \caption{\label{fig1}Snapshots of particles at the beginning ($t = 0$) and the end ($t = 4$ s) of the simulation with $W$ = 60$d_p$ ($H$ = 240$d_p$). Particles with initial random spatial distribution develop heterogeneities in particle concentration (better viewed on a 20$d_p$ thick slice). The enlarged view shows the detected agglomerates from a square region marked on the slice.}
 \end{figure}

Following our recent work \cite{Liu2017prf}, unbounded fluidization \cite{Fullmer2017} is simulated in a fully periodic domain with a square cross-section (Fig.~\ref{fig1}) via coupled computational fluid dynamics and discrete element method (CFD-DEM). Compared with no-slip side walls, the periodic domain removes bulk shear in the mean flow, thereby isolating the mechanism of deagglomeration associated with clusters. In CFD-DEM, particle trajectories are integrated via Newton’s equations of motion, where the contact forces between particles are related to particle overlap \cite{Cundall1979, Brilliantov1996, Poschel2005, Antypov2011}. For computational simplicity and convenient control of cohesion level, a constant cohesion is applied during physical contact of particles (i.e. ``square-force'' cohesion model with zero cut-off distance \cite{Liu2018}), following previous studies \cite{Waitukaitis2011, Kobayashi2013}. We recently demonstrated \cite{Liu2018} that this square-force cohesion model is a valid surrogate of more rigorous models where cohesion may depend on interparticle separation, surface morphologies, etc. \cite{Hamaker1937, Johnson1971, Derjaguin1975, Rumpf1990, Rabinovich2000a, Rabinovich2000b} The gas phase governed by the Navier-Stokes equations is solved using a cell size equal to two particle diameters \cite{Tsuji1993, Xu1997, vanderHoef2008}. The gas and solid phases are coupled via a local, solid-concentration-dependent drag law established from direct numerical simulations \cite{Hill2001no2, Hill2001no1, Benyahia2006}. The open-source solver MFiX \cite{Syamlal1993} is used to perform the simulations. Details on the numerical method are available elsewhere \cite{Liu2017prf}. In the simulations, the incompressible gas has density $\rho_g = 0.97$ kg$/$m$^3$ and viscosity $\mu_g$ = $1.8335\times10^{-5}$ Pa$\cdot$s. Particles are frictionless solid spheres with diameter $d_p = 69\times10^{-6}$ m, density $\rho_p = 2500$  kg$/$m$^3$, restitution coefficient $e$ = 0.97, Young's modulus $E = 10$ MPa and Poisson's ratio $\nu = 0.22$. Except where noted, a cohesion force $F_c = 680$ nN is applied. To study the effect of system size, we vary the domain width $W$ and height $H$ in proportion, with constant aspect ratio $\alpha = H/W \equiv 4$ (see Fig.~\ref{fig3}d). In simulations with $W$ and $H$ independently varied, flow properties show larger sensitivity to increasing $W$ than $H$, which is associated with flow anisotropy (see Supplemental Materials). The overall solid concentration $\epsilon_s$ = 0.01, corresponding to particle number count $N_p$ from 2,062 to 101,680 as $W$ varies from $30d_p$ to $110d_p$ ($H$ from $120d_p$ to $440d_p$). Particles are initially at rest and randomly placed throughout the domain. Gas flows in the upward direction ($y$-direction in Fig.~\ref{fig1}) at a constant superficial velocity $U =$ 43 cm/s. As time evolves, particles accelerate until they reach the terminal velocity, or statistical steady state. Our following analysis focuses on steady-state properties, i.e. time-averaged data over 1-4 s. (Note that varying $U$ does not affect the steady-state gas-solid slip velocity or the levels of clustering and agglomeration, see Supplemental Materials.)

It is worth noting that the flow regime examined here corresponds to that in typical risers \cite{Agrawal2001}, with particle Reynolds numbers ${{\mathop{\rm Re}\nolimits} _p} = {\rho _g}{d_p}{v_t}/{\mu _g} = 1.3$ and mean-flow Stokes number ${\rm{S}}{{\rm{t}}_M} = {\rho _p}{d_p}{v_t}/(9{\mu _g}) = 369.5$, where $v_t=\rho_pgd_p^2/(18\mu_g)$ is the particle terminal velocity in undisturbed fluid flow \cite{Garzo2012, Tenneti2014}. Therefore, fluid inertia and viscosity play a secondary role to particle inertia such that the flow, and more specifically agglomeration and breakage, is characterized by solid collisions \cite{Koch2001, Fullmer2017}. The current system therefore differs from common liquid-solid suspensions with much lower Stokes numbers (${\rm{S}}{{\rm{t}}_M} \sim O(1)$) \cite{Caflisch1985, Nicolai1995, Segre1997, Yin2008}, where deagglomeration in dilute suspensions is largely due to the solid-liquid interactions \cite{Lick1988, Burban1989, Stolzenbach1994, Winterwerp1998, Higashitani2001, Grabowski2011, Babler2015, Watanabe2017, Njobuenwu2018}. Examples of such low-Stokes systems are cohesive sediment transport \cite{Winterwerp2004, Burchard2018}. Furthermore, the effect of cluster-induced turbulence, which refers to the generation of gas-phase turbulence due to coupling with the solid phase (\cite{Capecelatro2014, Capecelatro2015, Capecelatro2016}), is light in our systems. Namely, the estimated ratio of turbulent viscosity associated with single-particle-induced turbulence (PIT) $\nu_{pit}$ to gas viscosity $\nu_{pit}/{\nu_g}$ (= 0.008) is much smaller than unity, where $\nu_{pit} = 0.6\epsilon_sd_pv_t$ \cite{Sato1981}.

We begin our discussion with solid-phase heterogeneities, quantified by a heterogeneity index $D$. This index characterizes the deviation of the particle-number-density fluctuation from that corresponding to a random distribution \cite{Fessler1994}; a larger $D$ indicates a higher level of heterogeneity. Specifically, $D = (\sigma-\sigma_p)/\mu$, where $\mu$ and $\sigma$ are the mean and standard deviation of the local number density, respectively. $\sigma_p$ is the standard deviation associated with the initial random placement of particles inside the domain, and $\sigma_p = [N_p/(l^3W^2H)]^{1/2}$, where $l = 10d_p$ is the cell size used in extracting the local particle number density \cite{Fessler1994}.

 \begin{figure}[t]
 \includegraphics[width=0.9\columnwidth]{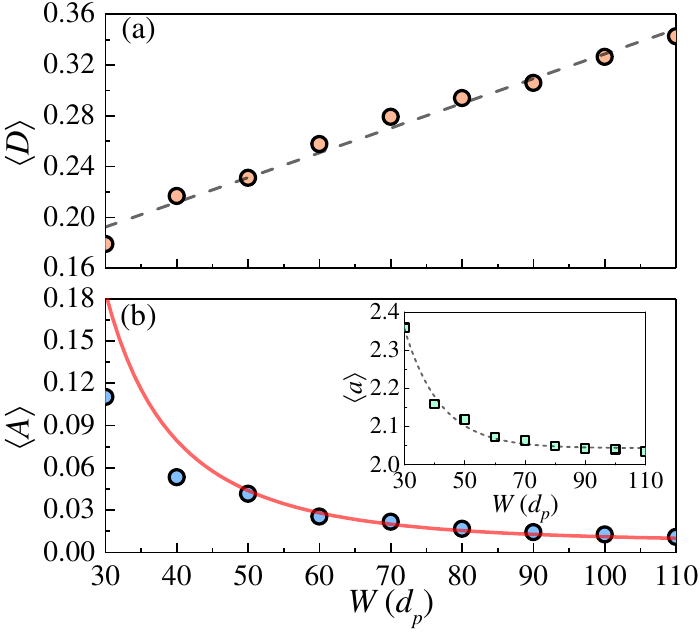}
 \caption{\label{fig2}Steady-state (a) heterogeneity index $\left\langle D \right\rangle$ and (b) fraction of particles in agglomerates $\left\langle A \right\rangle$ with increasing system size ($W$). Inset for (b): steady-state mean agglomerate size $\left\langle a \right\rangle$ with increasing $W$. Dashed lines are linear and exponential fits for (a) and the inset of (b), respectively, to guide the eye. Solid line on (b) is  Eq.~(\ref{eq1}); see text for details.}
 \end{figure}
 
 \begin{figure*}[t]
 \includegraphics[width=2.0\columnwidth]{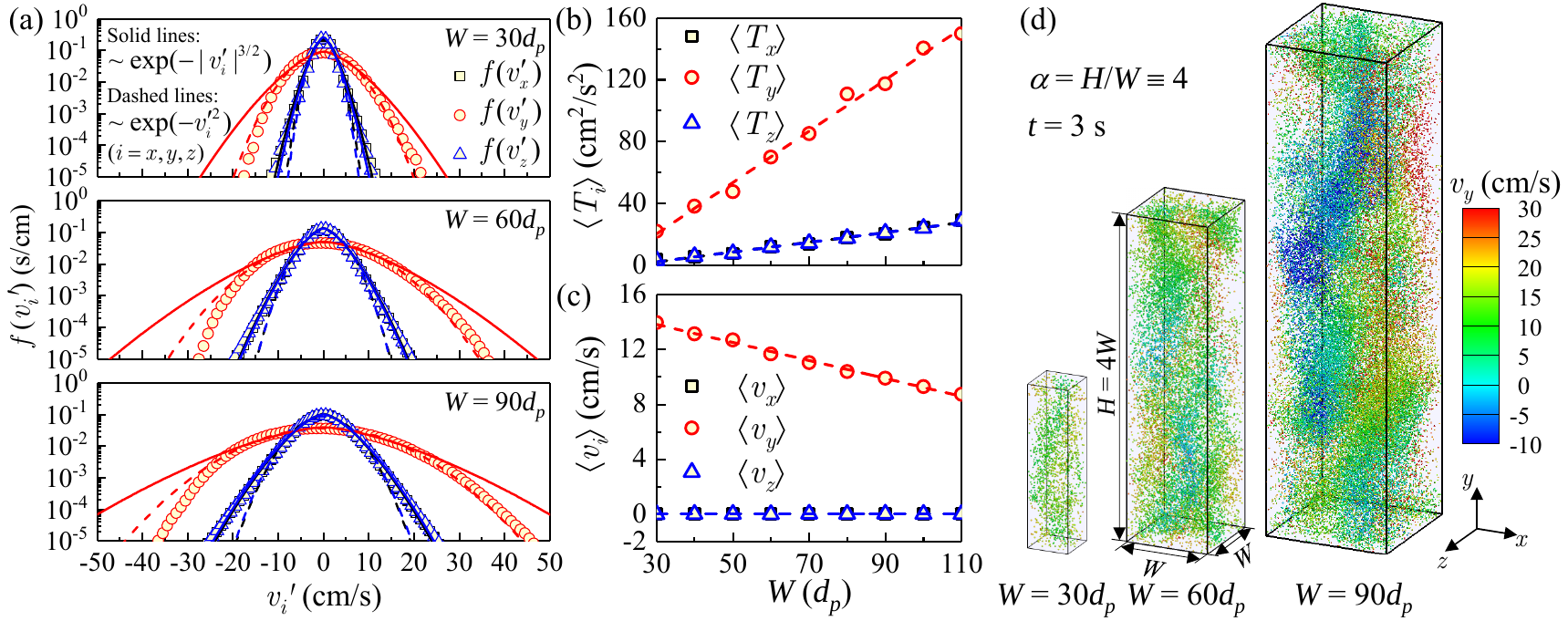}
 \caption{\label{fig3}(a) Steady-state distributions of particle fluctuating velocities in transverse $v'_x$, $v'_z$ and streamwise $v'_y$  directions with increasing $W$ (symbols). Dashed lines are Gaussian $f({v'_i}) = \sqrt {1/(2\pi \left\langle {{T_i}} \right\rangle )} \exp [ - {v'_i}^2/(2\left\langle {{T_i}} \right\rangle )],(i = x,y,z)$ and solid lines are fits of $f({v'_i}) = {A_1}\exp [ - |{v'_i}/({A_2}{\left\langle {{T_i}} \right\rangle ^{1/2}}){|^{3/2}}]$, where $A_1$ and $A_2$ are fitting parameters. Steady-state (b) granular temperatures $\left\langle T_i \right\rangle$ and (c) particle velocities $\left\langle v_i \right\rangle$ in three directions with increasing $W$. Dashed lines are linear fits to guide the eye. (d) Snapshots of particles with increasing $W$ at $t = 3$ s. See corresponding movies in Supplemental Materials.}
 \end{figure*}
 
Fig.~\ref{fig2}a shows the steady-state heterogeneity index $\left\langle D \right\rangle$ increases linearly with system size $W$ (while $\alpha \equiv 4$). Similar trends are reported for non-cohesive particles in granular \cite{Liss2001, Lasinski2004, Mitrano2011} and gas-solid flows \cite{Xiong2010, Capecelatro2016}, where the higher level of heterogeneities in larger systems is explained by the increased space for hydrodynamic instabilities to develop \cite{Liss2001, Garzo2005}, analogous to the laminar-to-turbulent transition in single-phase pipe flows. For cohesive particles, in addition to increased clustering, an increased $\left\langle D \right\rangle$ can also result from enhanced agglomeration. However, Fig.~\ref{fig2}b shows the levels of agglomeration decreases with $W$, in terms of both the steady-state fraction of particles in agglomerates $\left\langle A \right\rangle$ and the agglomerate size $\left\langle a \right\rangle$ (average number of particles in each agglomerate). To obtain $A$ and $a$, we isolate agglomerates from the overall system heterogeneity by tracking enduring contacts between particles. We associate particles with an agglomerate when their contact duration $t_c$ exceeds a critical value $t_{c,crit} = 59 \times10^{-6}$ s, considerably longer than the typical contact durations for non-agglomerating collisions \cite{Kellogg2017}. Agglomerate breakage is recorded when particles lose physical contact with the agglomerate/particle. In all systems, agglomerates are dominated by doublets ($a = 2$) ($\left\langle a \right\rangle < 2.4$ on inset of Fig.~\ref{fig2}b), consistent with Fig.~\ref{fig1}, where we zoom in on the flow pattern at $t = 4$ s to find a few doublets and only one triplet ($a = 3$). Since agglomeration decreases as $W$ increases, the increase in $\left\langle D \right\rangle$ with $W$ can only be attributed to an increased clustering. Thus, in contrast to granular flows where the reduced impact velocities within clusters enhance agglomeration, clustering in gas-solid flows appears to inhibit agglomeration. To probe the mechanism for this counterintuitive behavior, the particle velocity distributions are examined next since they dictate whether or not agglomeration occurs \cite{Thornton1991, Donahue2010}.

Fig.~\ref{fig3}a shows steady-state distributions for the three components $v'_i(i = x,y,z)$ of particle fluctuating velocities $\mathbf{v'} = \mathbf{v} - \bar{\mathbf{v}}$, where $\mathbf{v}$ and $\bar{\mathbf{v}}$  are instantaneous and mean particle velocities, respectively. In transverse directions, the distributions $f({v'_x})$ and $f({v'_z})$ deviate from Gaussian (dashed lines) and exhibit overpopulated tails $\sim\exp ( - |{v'_i}{|^{3/2}})$ (solid lines), which is the signature of driven, non-cohesive granular gases identified theoretically \cite{Puglisi1998, vanNoije1998}, numerically \cite{Moon2004, vanZon2005} and experimentally \cite{Losert1999, Olafsen1999, Rouyer2000, Blair2003no1, Reis2007, Scholz2017}. The consistency with granular systems is reasonable since particle-particle interactions dominate the dynamics in the transverse directions with zero mean flow ($\left\langle v_x \right\rangle$, $\left\langle v_z \right\rangle = 0$ on Fig.~\ref{fig3}c). On the other hand, the streamwise distributions $f({v'_y})$ are flatter and better described by Gaussians with positive skewness, which is attributed to the stronger gas-solid interactions (larger input of granular energy \cite{Warr1995, Grossman1997}) in the streamwise direction \cite{Ye2004, Ma2006, Capecelatro2015, Liu2017aichej, Vaidheeswaran2017}. As $W$ increases, the distributions get wider in all directions. Accordingly, the steady-state granular temperatures $\left\langle T_i \right\rangle$, defined as the variances of the three components of particle fluctuating velocity \cite{Jenkins1983}, increase with $W$ (Fig.~\ref{fig3}b), consistent with gas-solid flows of non-cohesive particles \cite{Capecelatro2016, Liu2017aichej}. The increased $\left\langle T_i \right\rangle$ with system size can be traced to increased clustering. The physical picture is that clusters tend to fall down as a result of ``jet-bypassing'' \cite{Capecelatro2015}: the gas bypasses clusters, leading to reduced drag, whereas an increased pressure drop is needed for the gas to squeeze through clusters (higher flow resistance in clusters). The falling clusters then collide with individual particles or small clusters/agglomerates entrained by the gas flowing upwards. These ``cluster-induced'' collisions provide an added source of granular energy, which increases with the clustering level and results in higher $\left\langle T_i \right\rangle$ in larger domains. In Fig.~\ref{fig3}d, falling clusters are increasingly visible with increasing $W$, i.e., more particles with lower or negative streamwise velocities $v_y$ are seen, leading to decreased $\left\langle v_y \right\rangle$ with $W$ (Fig.~\ref{fig3}c).

\begin{figure}[b]
 \includegraphics[width=0.9\columnwidth]{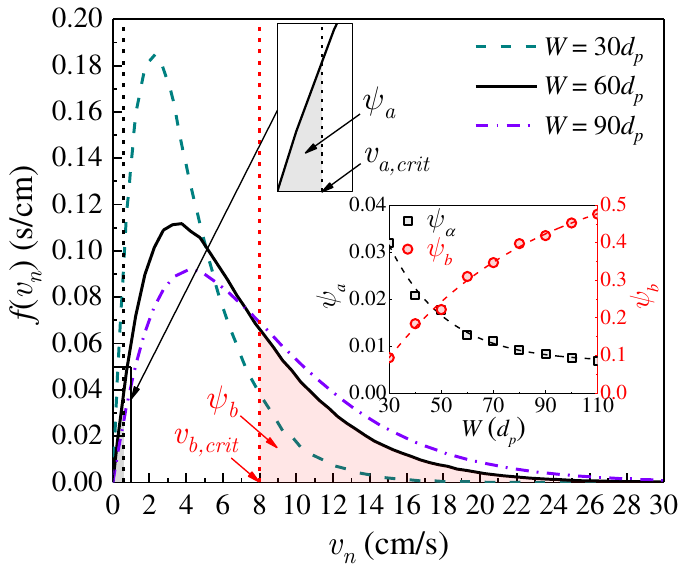}
 \caption{\label{fig4}Distributions of normal impact velocities $v_n$ for collisions collected during steady states with increasing $W$. The vertical dashed lines mark the critical agglomeration $v_{a,crit}$ and breakage velocities $v_{b,crit}$. The shaded areas demarcate the probabilities of agglomeration (black) and breakage (red) at $W = 60d_p$. Inset: $\psi_a$ and $\psi_b$ with increasing $W$, where dashed lines are exponential fits to guide the eye.}
 \end{figure}

Due to the increased $\left\langle T_i \right\rangle$, both the impact velocity and frequency of collisions increase, analogous to molecular gases at elevated thermal temperatures. Correspondingly, as shown in Fig.~\ref{fig4}, the steady-state distributions of the normal impact velocities $v_n$ (magnitude of the normal relative velocity right before a collision) shift to higher values with increasing $W$. Since agglomeration occurs at lower impact velocities \cite{Thornton1991, Donahue2010}, the increased $v_n$ is responsible for the decreasing agglomeration shown in Fig.~\ref{fig2}b.

To explain ``cluster-induced deagglomeration'' with more mathematical rigor, we propose an analytical model to relate $\left\langle A \right\rangle$ to the impact velocity distribution $f(v_n)$. First, recall in current system with light cohesion, the agglomerates are largely doublets (Fig.~\ref{fig1}). Thus, at statistical steady state, the generation and breakage rates of doublet are assumed equal, such that $\omega_{11}\psi_a = \omega_{12}\psi_b$, where $\omega_{11}$ and $\omega_{12}$ are, respectively, the frequencies of singlet-singlet and singlet-doublet collisions. $\psi_a$ and $\psi_b$ are, respectively, the probabilities (success factors \cite{Fan2004}) of collisions resulting in agglomeration of singlets (from singlet-singlet collisions) and breakage of doublets (from singlet-doublet collisions). The collision frequencies ${\omega _{11}} = 1/2n_1^2{g_0}{s_{11}}\langle {v_{rel}}\rangle$ and ${\omega _{12}} = {n_1}{n_2}{g_0}{s_{12}}\langle {v_{rel}}\rangle$ \cite{Gidaspow1994, Kumaran2009}, where $n_1$ and $n_2$ are number densities of singlets and doublets, $g_0$ is the radial distribution function at contact, $s_{11}$ and $s_{12}$ are, respectively, the collisional cross section areas for singlet-singlet and singlet-doublet collisions, and $\left\langle v_{rel} \right\rangle$ is the mean particle relative velocity magnitude. Since $n_1 = N_p(1- \left\langle A \right\rangle)/V$, $n_2 = N_p\left\langle A \right\rangle/(2V)$, where $V$ is the system volume, combining the above relations gives
\begin{equation} \label{eq1}
\left\langle A \right\rangle  = {\left( {\frac{{{s_{12}}{\psi _b}}}{{{s_{11}}{\psi _a}}} + 1} \right)^{ - 1}}.
\end{equation} 
For singlet-singlet collisions, $s_{11}=\pi d_p^2$. The singlet-doublet collisional cross section $s_{12}$ depends on the orientation of doublets. In Supplemental Materials, we derive the average collisional cross section over possible doublet orientations in the current system, and obtain  $s_{12}=4.66 d_p^2$. Next, we evaluate $\psi_a$ and $\psi_b$. For singlet-singlet collisions, agglomeration occurs when $v_n$ is below the critical agglomeration velocity $v_{a,crit}$. Using a dimensional analysis, we recently \cite{Liu2017cej} derived an expression relating $v_{a,crit}$ to particle material properties: $v_{a,crit} = cF_c^{5/6}{(1 - {\nu ^2})^{1/3}}/(d_p^{5/3}\rho _p^{1/2}{E^{1/3}})$, where $c$ is a dimensionless parameter dependent on particle restitution coefficient $e$. In this work, $e = 0.97$ and $c = 1.042$ \cite{Liu2017cej}, giving $v_{a,crit} = 0.6$ cm/s. In this same work \cite{Liu2017cej}, we conducted controlled simulations of singlet-doublet collisions for particles used here. We found the critical breakage velocity $v_{b,crit}$ (i.e. when $v_n > v_{b,crit}$, the doublet breaks and the collision results in three singlets) depends on the relative position of the singlet and doublet before colliding (pre-collisional configurations). For simplification in the current analytical model, we use $v_{b,crit} = 8.0$ cm/s, which is the averaged $v_{b,crit}$ collected in controlled simulations sweeping all possible pre-collisional configurations \cite{Liu2017cej}. Therefore, we compute ${\psi _a} = \int_0^{{v_{a,crit}}} {f({v_n})d{v_n}}$ and ${\psi _b} = \int_{{v_{b,crit}}}^\infty  {f({v_n})d{v_n}}$. For example, $\psi_a$ and $\psi_b$ corresponding to $f(v_n)$ at $W = 60d_p$ are marked on Fig.~\ref{fig4} as shaded areas. As $W$ increases, $f(v_n)$ shifts towards higher values, causing $\psi_a$ to decrease and $\psi_b$ to increase (inset of Fig.~\ref{fig4}). Plugging $s_{11}$, $s_{12}$, $\psi_a$ and $\psi_b$ in Eq.~(\ref{eq1}), we find the decreasing $\left\langle A \right\rangle$ with increasing $W$ is well captured by Eq.~(\ref{eq1}) (solid line on Fig.~\ref{fig2}b). Quantitative agreement is observed except at $W \leq 40d_p$, possibly due to the increasing number of agglomerates larger than doublets (more rapid growth of $\left\langle a \right\rangle$ when $W \leq 40d_p$ on inset of Fig.~\ref{fig2}b), which are not considered in current model.

\begin{figure}[t]
 \includegraphics[width=0.9\columnwidth]{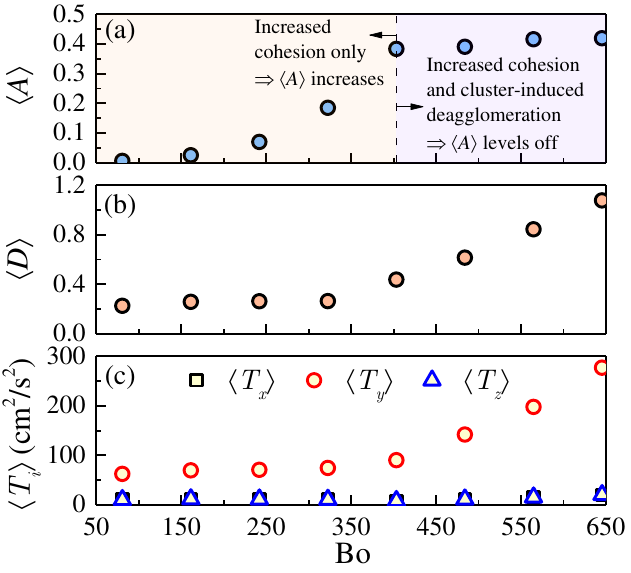}
 \caption{\label{fig5}Steady-state (a) agglomerate fraction $\left\langle A \right\rangle$, (b) heterogeneity index $\left\langle D \right\rangle$ and (c) granular temperatures $\left\langle T_i \right\rangle$ as a function of granular bond number Bo = $F_c/mg$ under $W \equiv 60d_p$ ($H \equiv 240d_p$).}
 \end{figure}

As additional evidence of cluster-induced deagglomeration beyond the lightly-cohesive systems ($F_c \equiv 680$ nN) examined thus far, we plot steady-state flow properties in Fig.~\ref{fig5} for systems with increasing granular Bond number Bo (Bo = $F_c/mg$, where $m$ is the mass of a single grain and $F_c$ varies from 340 nN to 2720 nN) for a fixed system size. When Bo $< 400$, $\left\langle T_i \right\rangle$ stays relatively constant so that $\left\langle A \right\rangle$ grows due to the increasing critical agglomeration and breakage velocities with increasing Bo. When  Bo $> 400$, the enhanced energy dissipation in collisions among particles with stronger cohesion leads to more prominent clustering as well as agglomeration ($\left\langle D \right\rangle$ increases evidently). However, the increasing level of clustering also triggers a rapid growth in $\left\langle T_i \right\rangle$, which contributes to deagglomeration. Consequently, instead of asymptotically approaching unity with increasing Bo as seen in gravity-free granular flows \cite{Weber2004}, $\left\langle A \right\rangle$ levels off at $\sim 0.4$ under the competing effects of increasing cohesion in gas-solid flows: i) increased agglomeration and (ii) increased cluster-induced deagglomeration.
 
In sum, an inverse response of clustering and agglomeration to increasing system size is identified in dilute gravity-driven gas-solid flows of lightly cohesive particles, which is explained by cluster-induced deagglomeration. Specifically, higher levels of clustering in larger systems enhance the relative velocities of particles, serving as a source of granular temperatures and higher collisional impact velocities that contribute to deagglomeration. The same mechanism explains the unexpected saturation of agglomeration levels as cohesion increases in gravity-driven gas-solid flows. Therefore, it is the gravity and the resulting increased collisional velocities between the falling (large) clusters and rising particles or (small) clusters that leads to the cluster-induced deagglomeration. Collectively, such interplay between clusters and agglomerates will impact numerous multiphase operations, where the interphase drag, heat transfer and chemical reactions rates are dependent on the nature of particle contacts (brief in clusters vs. enduring in agglomerates, etc.) \cite{Beetstra2006, Kosinski2013, Shah2013, Mehrabadi2016}. The identification of the cluster-induced deagglomeration warrants its consideration in related population balance efforts \cite{Kim2002, Kantak2009, Murphy2015, Kellogg2017} for developing continuum models of cohesive particles. Beyond gas-solid flows, the findings may have ramifications for (high Stokes number) gravity-driven liquid-solid suspensions \cite{Guazzelli2011}, colloids \cite{Barros2014}, emulsions and foams \cite{Nagel2017}, where hydrodynamic instabilities and long-range interparticle attractions coexist.

\begin{acknowledgments}
The authors are grateful for the financial support provided by the Dow Corning Corporation, a wholly owned subsidiary of the Dow Chemical Company, and National Science Foundation (CBET-1707046). The authors thank Kevin Kellogg and Casey LaMarche for stimulating discussions. This work utilized the RMACC Summit supercomputer, which is supported by the National Science Foundation (awards ACI-1532235 and ACI-1532236), the University of Colorado Boulder, and Colorado State University. The Summit supercomputer is a joint effort of the University of Colorado Boulder and Colorado State University.
\end{acknowledgments}

%

\end{document}